\documentclass[]{spie}  

\usepackage{amsmath,amsfonts,amssymb}
\usepackage{graphicx}
\usepackage[colorlinks=true, allcolors=blue]{hyperref}

\title{Somatomotor-Visual Resting State Functional Connectivity Increases After Two Years in the UK Biobank Longitudinal Cohort}
\author[a]{Anton Orlichenko}
\author[b]{Kuan-Jui Su}
\author[b]{Qing Tian}
\author[b]{Hui Shen}
\author[b]{Hong-Wen Deng}
\author[a]{Yu-Ping Wang}

\affil[a]{Department of Biomedical Engineering, Tulane University, New Orleans, LA 70118}
\affil[b]{Center for Biomedical Informatics and Genomics, School of Medicine, Tulane University, New Orleans, LA 70118}

\authorinfo{Further author information: (Send correspondence to Anton Orlichenko)\\Anton Orlichenko: E-mail: aorlichenko@tulane.edu}

\begin{document}
\maketitle
 
\begin{abstract}
\textbf{Purpose:} Functional magnetic resonance imaging (fMRI) and functional connectivity (FC) have been used to follow aging in both children and older adults. Robust changes have been observed in children, where high connectivity among all brain regions changes to a more modular structure with maturation. In this work, we examine changes in FC in older adults after two years of aging in the UK Biobank longitudinal cohort. \textbf{Approach:} We process data using the Power264 atlas, then test whether FC changes in the 2,722-subject longitudinal cohort are statistically significant using a Bonferroni-corrected t-test. We also compare the ability of Power264 and UKB-provided, ICA-based FC to determine which of a longitudinal scan pair is older. \textbf{Results:} We find a 6.8\% average increase in SMT-VIS connectivity from younger to older scan (from $\rho=0.39$ to $\rho=0.42$) that occurs in male, female, older subject ($>65$ years old), and younger subject ($<55$ years old) groups. Among all inter-network connections, this average SMT-VIS connectivity is the best predictor of relative scan age, accurately predicting which scan is older 57\% of the time. Using the full FC and a training set of 2,000 subjects, one is able to predict which scan is older 82.5\% of the time using either the full Power264 FC or the UKB-provided ICA-based FC. \textbf{Conclusions:} We conclude that SMT-VIS connectivity increases in the longitudinal cohort, while resting state FC increases generally with age in the cross-sectional cohort. However, we consider the possibility of a change in resting state scanner task between UKB longitudinal data acquisitions.
\end{abstract}


\section{Introduction}

Functional magnetic resonance imaging (fMRI) is a non-invasive technique that has proven indispensable for investigating human neural processes in vivo.\cite{Belliveau1991FunctionalMO} For example, it has been used to localize the areas associated with vision,\cite{Cox2003-tu} attention,\cite{Coull1998-op}\cite{Pugh1996-tm} emotion,\cite{Phan2002-nb}\cite{6789831}\cite{Koelsch2006-ci} and language\cite{Hernandez2001-ec} to specific regions in the cortex, or at least find the regions that are most significantly involved in a specific task. Functional connectivity (FC) is a quantity derived from fMRI that measures the time correlation of blood oxygen level-dependent (BOLD) signal between different regions in the brain.\cite{Van_den_Heuvel2010-ig} FC has recently been used to predict age,\cite{10002422}\cite{8666981} sex,\cite{ICER2020105444}\cite{9146335} race,\cite{Orlichenko2023-xl} psychiatric disease status,\cite{Wang2018-iv}\cite{Rashid2016-ct} and pre-clinical Alzheimer's disease.\cite{MILLAR2022119228} Efforts to predict general fluid intelligence, although common,\cite{Qu2021EnsembleMR}\cite{9146335} are thought by some to be confounded by differential achievement score distribution among ethnicities and the strong presence of race signal in FC.\cite{Orlichenko2023-xl} FC has proven effective in predictive studies because of its simplicity and its robust representation of complex BOLD signal activity, as evidenced by high subject identifiability across different scanner tasks and across time.\cite{Finn2015-ft}\cite{Cai2019-vp}\cite{Orlichenko2023-ad}

Besides being used as a predictive tool, FC has been observed to undergo changes throughout the lifespan. For example, connectivity in young children is generally very high between all brain regions and decreases while also becoming more modularized during and after puberty.\cite{Dosenbach2010-ef} The FC of males and females is also quantitatively different, with females having higher intra-DMN connectivity and males having relatively greater connectivity between the DMN and other networks, although there is a wide degree of individual variation.\cite{Mak2017-yj}\cite{Ficek-Tani2022-yw} Meanwhile, studies have shown that changes occur in the DMN during late middle and old age,\cite{Fjell2017-av} although the exact direction of change in FC does not always appear constant.\cite{Staffaroni2018-kq} In addition, various studies have examined age-related changes in the cingulum\cite{Hirsiger2016-jo} and medial temporal lobe.\cite{Salami2016-pz} Given the recent interest in using fMRI to predict pre-clinical Alzheimer's disease,\cite{Staffaroni2018-kq}\cite{MILLAR2022119228} we believe a knowledge of ordinary changes in FC during old age is essential. This is especially true because it has been shown that a confounder can easily be mistaken for a true signal indicative of, e.g., general fluid intelligence or achievement score.\cite{Orlichenko2023-xl}

This study uses the longitudinal cohort of the UKB\cite{Sudlow2015-zq} to examine changes in the FC of individuals after an average of two years, the time between longitudinal scans. The UKB population of subjects with fMRI scans is predominantly (98\%) Caucasian, ruling out race as a possible confounding effect. Additionally, we investigate changes in FC in longitudinal sub-populations based on subject age and sex. We find that average FC between SMT-VIS networks increases on average from the first scan to the second, and that SMT and VIS-related connectivities are more predictive of scan age than those of other networks. The complete FC, or a large subset, is still required to attain the best accuracy.

\section{Methods}

We first describe the UKB dataset and the longitudinal subset used for our analysis. We then describe pre-processing of the fMRI data and conversion into FC. Finally, we discuss prediction of older vs younger scan in the longitudinal cohort and detail our methods for analysis of FC changes.

\subsection{UK Biobank Longitudinal Cohort}

The UKB contains various data of more than 500,000 subjects in the UK, of who more than 40,000 have fMRI scans.\cite{Sudlow2015-zq} We processed two longitudinal resting state scans for 2,722 subjects, taken approximately two years apart. These subjects are approximately equally split between male and female, and have significant numbers of younger and older adults. The longitudinal cohort is composed of 1,289 genetic males and 1,369 genetic females, with the rest not having genetic sex information. The ethnicity of the subset of the UKB with fMRI scans is 98\% Caucasian. Besides the 2,722 subjects we processed, an additional 154 subjects have the second longitudinal scan but not the first, resulting either from missing original source data or a failure in our SPM12-based preprocessing pipeline.

\subsection{fMRI Preprocessing}
\label{sec:preproc}

\begin{figure}
    \centering
    \includegraphics[width=17cm]{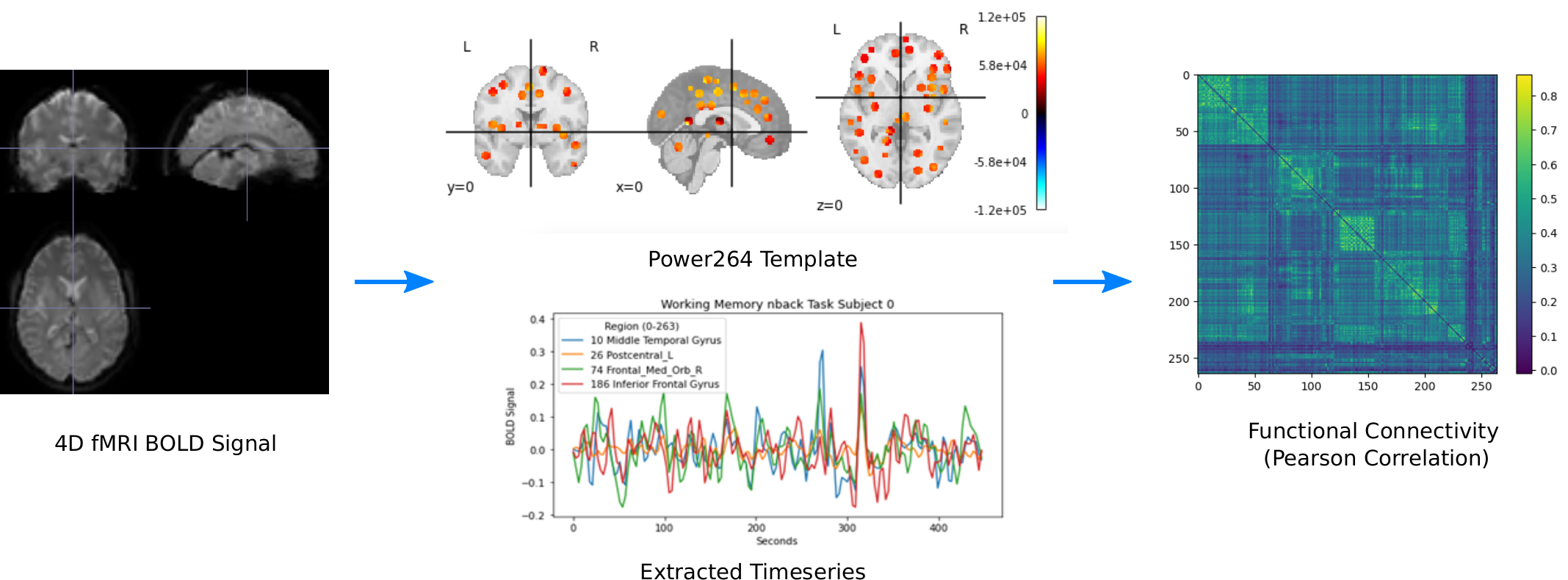}
    \caption{Preprocessing pipeline for converting 4D fMRI volumes into FC using the Power264 atlas.\cite{Power2011-fd} Reproduced with permission from Orlichenko et al. (2023).\cite{Orlichenko2023}}
    \label{fig:pipeline}
\end{figure}

The original scan acquisition parameters are described elsewhere,\cite{UKB-Brain-Imaging-Doc}\cite{Alfaro-Almagro2018-be} but consist of both resting state and task fMRI scans with a repetition time of $\text{TR}=0.735$ sec. For this study, we examined the resting state scans only. All resting state 4D fMRI volumes were processed with SPM12, including co-registration and warping to MNI space (\url{http://www.fil.ion.ucl.ac.uk/spm/software/spm12/}). BOLD signal was extracted using the Power264 atlas,\cite{Power2011-fd} which consists of 264 ROIs grouped into 14 functional networks and represented by 5mm radius spheres. The resulting timeseries were bandpass filtered between 0.01 and 0.15 Hz to remove scanner drift, noise, heartbeat, and some breathing signal. Pearson correlation of the filtered timeseries was used to create subject-specific FC matrices, which were reduced to the unique entries in the upper right triangle and vectorized. The entire procedure is summarized in Figure~\ref{fig:pipeline}.

In contrast to the Power264 atlas-derived FC constructed by us, the original UKB data provided the unique part of 21-region and 55-region FC and partial correlation-based connectivity (PC) matrices based on ICA in vectorized format.\cite{UKB-Brain-Imaging-Doc} These matrices were calculated through the use of PCA on whole cohort fMRI data followed by ICA,\cite{UKB-Brain-Imaging-Doc} meaning that regions overlap in an unpredictable way and are not associated with specific functional networks. Although prediction using 55-component ICA-based FC and PC is often as good as and sometimes better than prediction using Power264 atlas-derived FC, the resulting connectivities are uninterpretable with regards to BOLD signal within specific regions. Additionally, in predicting which scan is older, Power264 asymptotes to a higher predictive accuracy than either of the ICA-derived measures (see Figure~\ref{fig:prediction}).

\subsection{Prediction of Scan Order and Analysis of FC}

Prediction of scan age in the UKB longitudinal cohort was carried out by logistic regression (\url{https://scikit-learn.org/stable/}) models with 20 bootstrapping repetitions, using the scikit-learn implementation.\cite{scikit-learn} The regularization parameter was fixed to $C=1$, which was found to be near the optimal value for all training set sizes using grid search. It was found that a simple difference of scan FCs gave the best prediction results compared to concatenation or difference and concatenation, using either logistic regression or MLP. The training set was created with randomization of whether older scan was subtracted from younger scan or younger scan was subtracted from older scan. Our code for computing prediction accuracy can be found online (\url{https://github.com/aorliche/ukb-longitudinal-smt-vis}). However, UKB data sharing policy precludes us from posting the longitudinal data itself; interested researchers may contact us with any questions.

Analysis of FC was performed by finding the mean (Figure~\ref{fig:main}) and standard deviation (Figure~\ref{fig:std}) of older scan FC minus younger scan FC for the longitudinal cohort. Additionally, prediction of scan order was carried out using the average connectivity between each of the Power264 networks, each network consisting of many individual ROIs. As before, logistic regression with 20 bootstrap repetitions and $C=1$ was used for this purpose. A Bonferroni-corrected two-sided t-test was applied to the 105 average inter-network connectivity differences (from the complete graph of 14 functional networks) of the 2,722 longitudinal subjects to determine if they were significantly different from zero (Figure~\ref{fig:pred-inter-net} Bottom).

\section{Results}

We first describe trends in FC changes during the average of 2 years between longitudinal scans, summarize the ability of simple machine learning models to identify older vs younger scan, and investigate the ability of specific inter-network connectivities to predict scan order. We then summarize the statistical significance of inter-network FC changes with aging, in both the longitudinal and cross-sectional cohorts of the UKB. Finally, we consider the possibility that the observed longitudinal changes are due to a change in scanner task by presenting inter-task FC differences in the Philadelphia Neurodevelopmental Cohort (PNC) dataset.\cite{Satterthwaite2014NeuroimagingOT}

\subsection{Inter-Network FC Changes}

\begin{figure}
    \centering
    \hspace*{-1cm}\includegraphics[width=18cm]{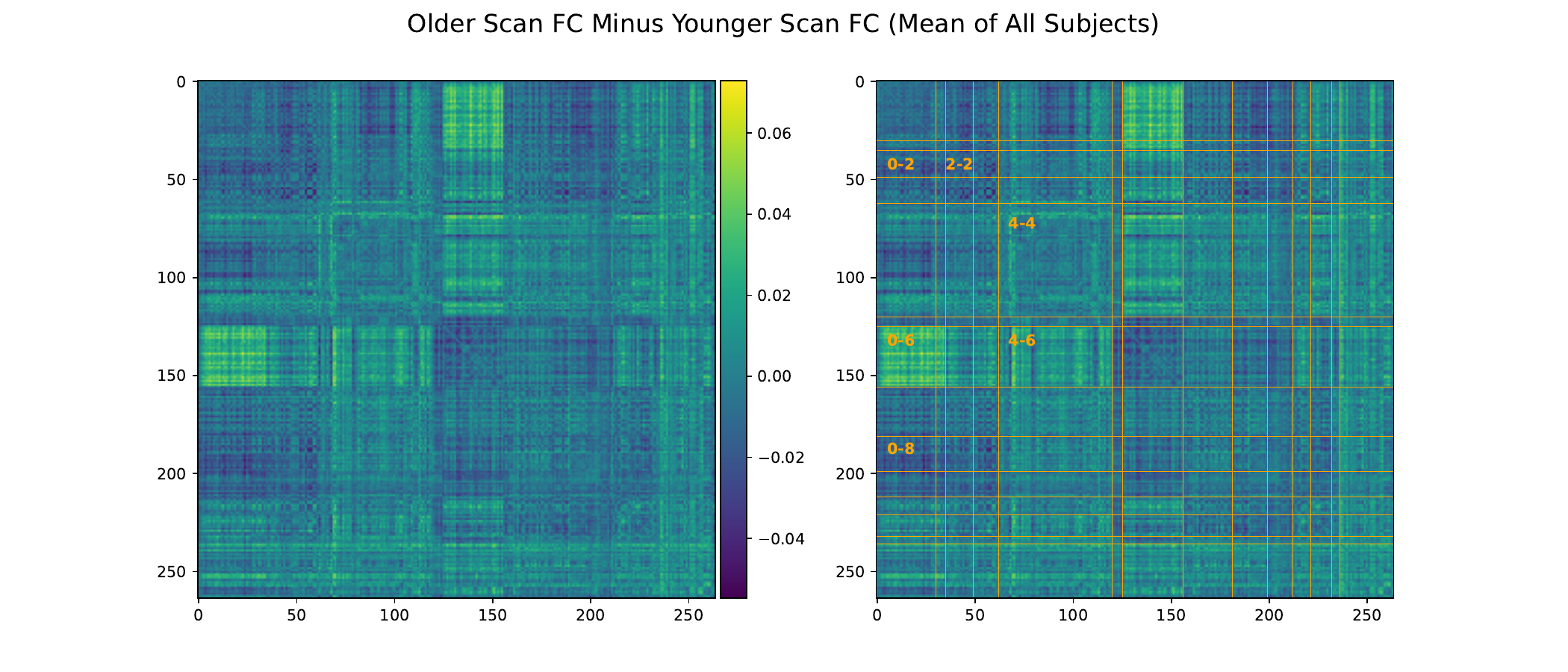}
    \caption{Difference in FC calculated by older scan minus younger scan, averaged over all 2,722 longitunidal cohort subjects. There are significant average differences in SMT-VIS connectivity (labeled 0-6). The same plot is displayed on the left and right, with Power264 network divisions on the right hand side. Network labels can be found in Table~\ref{tab:regions}.}
    \label{fig:main}
\end{figure}

In Figure~\ref{fig:main}, we show that, on average, SMT-VIS connectivity increases from younger scan to older scan. The right hand side of Figure~\ref{fig:main} displays divisions of the 14 functional networks included in the Power264 atlas. Network labels and abbreviations are listed in Table~\ref{tab:regions}. The increase in connectivity is large and distinct over the majority of SMT-VIS FCs compared to other non-SMT and non-VIS FCs. Many FCs involving the VIS network appear to increase in connectivity from the first scan to the second. The average change in FC in the SMT-VIS connection is $6.8\%$, corresponding to a mean change $\mu_{\Delta\rho}=+0.03$, compared to a standard deviation of $\sigma_{\Delta\rho}=0.26$. Figure~\ref{fig:pred-inter-net} shows that, although small compared to the standard deviation, this difference is very significant.

\begin{figure}
    \centering
    \hspace*{-1cm}\includegraphics[width=18cm]{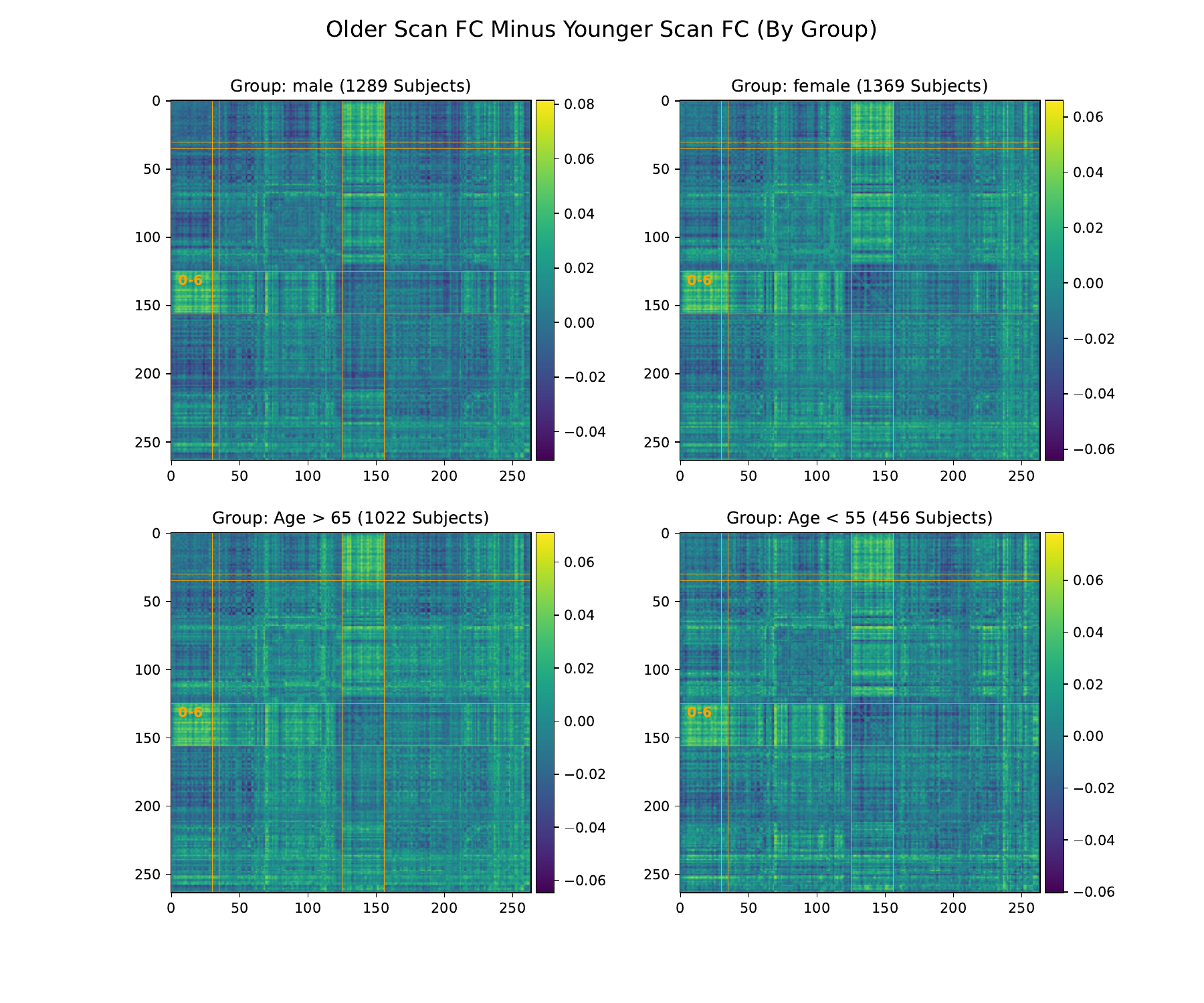}
    \caption{Significant increase in SMT-VIS connectivity after an average of 2 years in the UKB longitudinal cohort appears in male, female, younger, and older groups, and seems to be an invariant feature of FC change in the longitudinal UKB cohort.}
    \label{fig:main-groups}
\end{figure}

Figure~\ref{fig:main-groups} displays the same analysis, i.e., the average change from first scan to second, for four subsets of the cohort. These subsets are male subjects, female subjects, young ($<55$ years old) subjects, and old ($>65$ years old) subjects. All four subsets observed the same effect as the whole cohort, thus we rule out very old age or gender as confounding factors.

\begin{table}
    \caption{Regions, abbreviations, and labels in the Power264 atlas.}
    \begin{center}
    \begin{tabular}{|c|c|l|c|c|l|}
        \multicolumn{6}{c}{\textbf{Functional Networks}} \\
        \hline
        Label & ROIs & & Label & ROIs & \\
        \hline
         0 & 0-29 & Somatomotor Hand (SMT) & 7 & 156-180 & Frontoparietal (FRNT) \\ 
         \hline
         1 & 30-34 & Somatomotor Mouth (SMT) & 8 & 181-198 & Salience (SAL) \\ \hline
         2 & 35-48 & Cinguloopercular (CNG) & 9 & 199-211 & Subcortical (SUB) \\ \hline
         3 & 49-61 & Auditory (AUD) & 10 & 212-220 & Ventral Attention (VTRL) \\ 
         \hline
         4 & 62-119 & Default Mode (DMN) & 11 & 221-231 & Dorsal Attention (DRSL) \\ \hline
         5 & 120-124 & Memory (MEM) & 12 & 232-235 & Cerebellar (CB) \\ \hline
         6 & 125-155 & Visual (VIS) & 13 & 236-263 & Uncertain (UNK) \\ \hline
    \end{tabular} \\
    \end{center}
    \label{tab:regions}
\end{table}

\subsection{Predicting Older Scan of Pair}

\begin{figure}
    \centering
    \includegraphics[width=12cm]{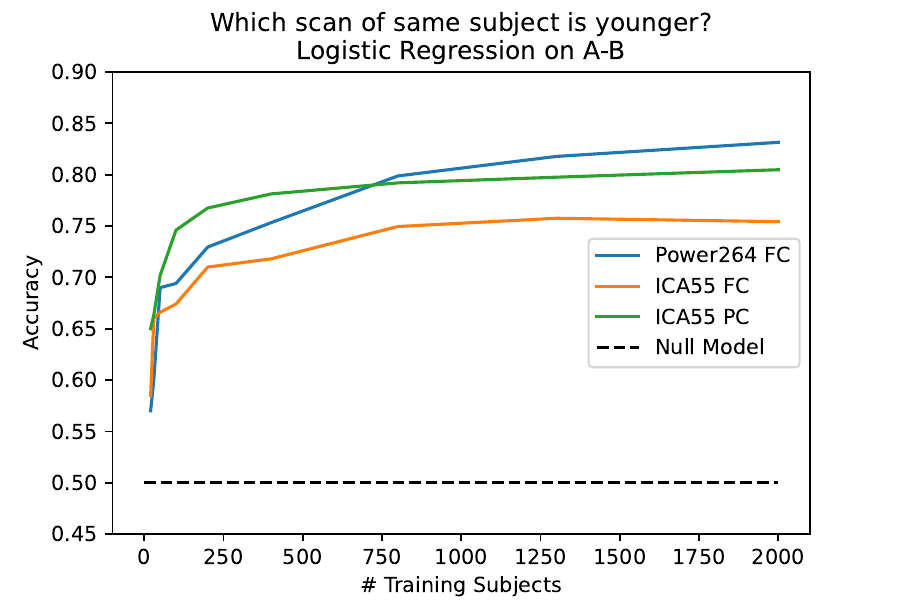}
    \caption{Capability of predicting older scan of a pair based on the difference of FC between the two scans, as a function of the number of training subjects. Three inputs are used: 55 component ICA FC, 55 component ICA PC, and the Power264 atlas FC. Prediction accuracy asymptotes at 82.5\%.}
    \label{fig:prediction}
\end{figure}

In Figure~\ref{fig:prediction}, one can see that is possible to predict which scan of a longitudinal pair is older with the Power264 atlas at an accuracy of 82.5\%, having 2,000 subjects in the training set and the rest in the test set. This measurement was repeated with 20 bootstrap iterations and averaged. The entire 34,716-feature upper right triangle of the FC matrix was used ot make the prediction. One can also see that the ICA FC/PC matrices provided pre-processed along with the UKB data are also able to predict scan order, although at a slightly reduced accuracy. Prediction is possible at an accuracy of 60-70\% using only 100-200 training set subjects.

\subsection{Prediction of Older Scan Using Specific Inter-Network Connections}

\begin{figure}
    \centering
    \includegraphics[width=16cm]{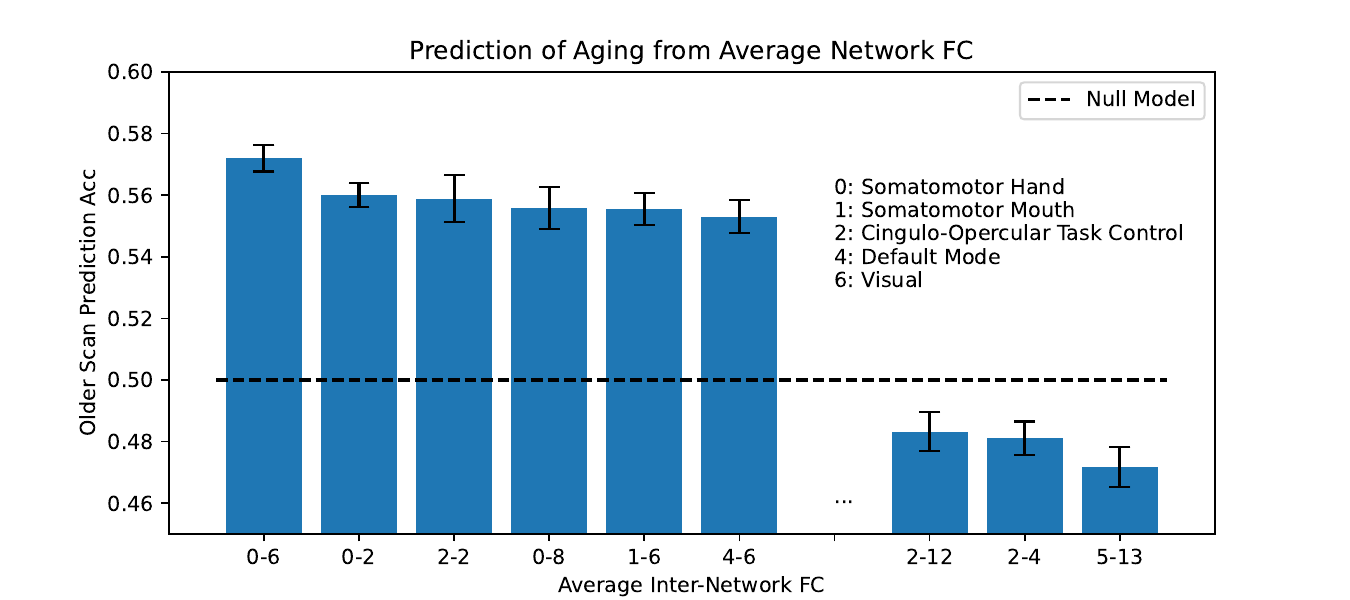} \\
    \includegraphics[width=11cm]{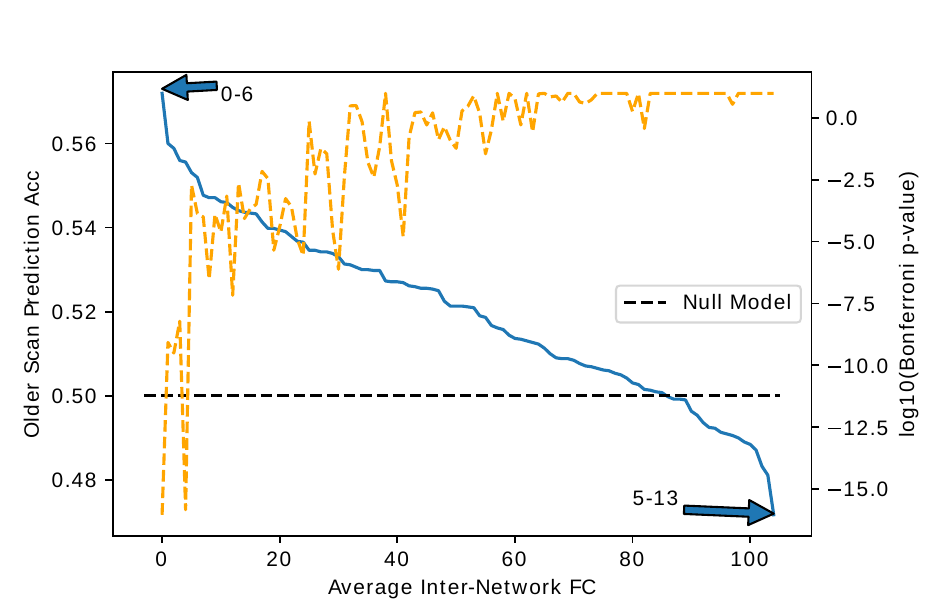}
    \caption{Ability to predict which scan of a subject is older based on average connectivity between regions. Top: best and worst inter-network connectivities for prediction. Bottom: Prediction accuracy for all 105 inter-network connectivities. We find that SMT-VIS connectivity has the maximum predictive ability of all regions at 57\%. In general, network-level connectivities involving SMT and VIS networks have higher predictive ability compared to other regions. The dashed orange line displays negative log base 10 of the Bonferroni-corrected p-value for significance of FC change between scans.}
    \label{fig:pred-inter-net}
\end{figure}

In Figure~\ref{fig:pred-inter-net}, we rank average inter-network FCs in their ability to predict scan order. As expected from the mean change in FC (Figure~\ref{fig:main}), the SMT-VIS connection is the most predictive of longitudinal scan age. Furthermore, SMT and VIS networks are included among the next several most predictive inter-network connections. In Figure~\ref{fig:pred-inter-net} bottom, we plot the predictive ability of all 105 inter-network connections, along with a p-value for the inter-network FC change being significantly different from zero. The raw p-value has been multiplied by 105 to account for multiple comparisons. It is highly significant for the first 10 or so most predictive inter-network connections.

Table~\ref{tab:incdec} lists the number of subjects whose FC increased or decreased for the SMT-VIS connection and over the entire brain. The table is divided among the four subsets of the longitudinal cohort mentioned previously. Additionally, we correlated several dozen subject phenotypes and longitudinally-tracked variables with changes in FC and report the most significant in Appendix~\ref{sec:corr}. In that section, we find an interesting but small correlation with hand grip strength, body mass index (BMI), and basal metabolic rate. In Section~\ref{sec:cross}, we find that average resting state FC increases with age across most inter-network connections in the much larger UKB cross-sectional cohort.

\begin{table}
    \label{tab:incdec}
    \caption{Number of subjects in the longitudinal cohort increasing and decreasing in average FC within the SMT-VIS connection and within the whole brain.}
    \begin{center}
    \begin{tabular}{|c|c|c|c|c|c|}
        \hline
         Group & +SMT-VIS FC & -SMT-VIS FC & +Total FC & -Total FC & Total Subjects \\
         \hline
         Male & 778 (60.4\%) & 511 & 690 (53.5\%) & 599 & 1289 \\
         \hline
         Female & 741 (54.1\%) & 628 & 671 (49.0\%) & 698 & 1369 \\
         \hline
         $<55$ years old & 269 (59.0\%) & 187 & 249 (54.6\%) & 207 & 456 \\
         \hline
         $>65$ years old & 577 (56.5\%) & 445 & 520 (50.1\%) & 502 & 1022 \\
         \hline
    \end{tabular}
    \end{center}
\end{table}

\begin{figure}
    \centering
    \includegraphics[width=16cm]{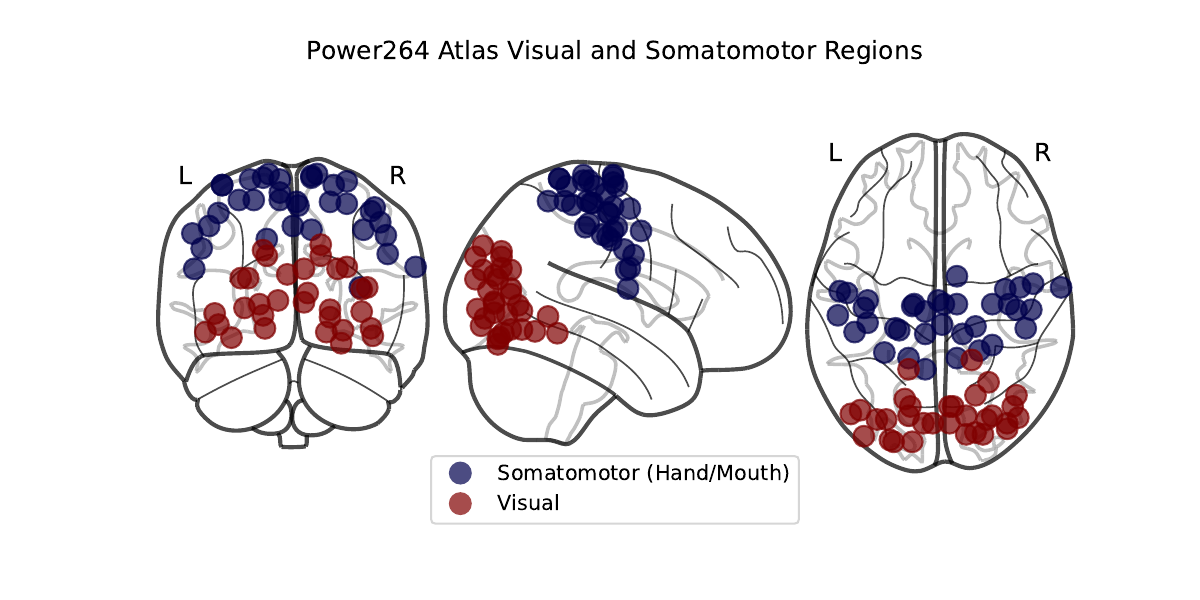}
    \caption{SMT and VIS regions in the Power264 atlas.}
    \label{fig:smt-vis-regions}
\end{figure}

\subsection{FC Changes with Age in the UKB Cross-Sectional Cohort}
\label{sec:cross}

\begin{figure}
    \centering
    \includegraphics[width=16cm]{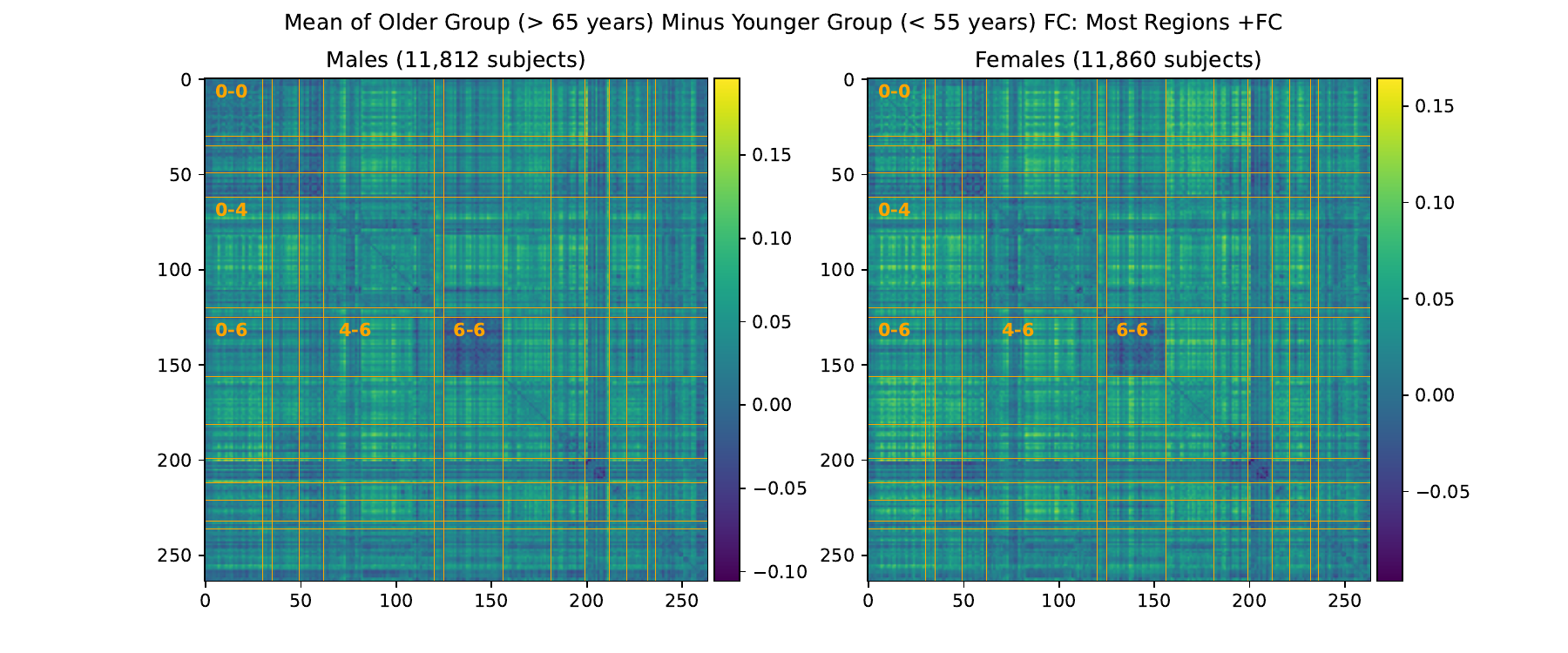}
    \caption{Mean FC change from younger group to older group in the large UKB cross-sectional cohort.}
    \label{fig:cross}
\end{figure}

We find that average resting state FC has a significant increase in almost all inter-network connections in the UKB cross-sectional cohort. Average maps of FC change are shown in Figure~\ref{fig:cross}. We fail to find a higher SMT-VIS change compared to other connections; however, almost all inter-network regions have a large positive increase in FC with aging. We give precise numbers for four inter-network connections as well as total FC in Table~\ref{tab:cross}.

\begin{table}
    \caption{Average FC changes with aging in the UKB cross-sectional cohort from young subjects ($<55$ years old) to old subjects ($>65$ years old).}
    \begin{center}
    \begin{tabular}{|c|c|c|c|c|c|c|}
        \hline
        & \multicolumn{3}{c|}{Male (Young to Old)} & \multicolumn{3}{c|}{Female (Young to Old)} \\
        \hline
         Regions & FC Increase & Std Dev of Avg FC & p-value & FC Increase & Std Dev of Avg FC & p-value \\
         \hline
         SMT-VIS (0-6) & 0.031 & 0.13 & $<10^{-23}$ & 0.029 & 0.13 & $<10^{-25}$ \\
         \hline
         SMT-DMN (0-4) & 0.045 & 0.11 & $<10^{-78}$ & 0.043 & 0.11 & $<10^{-82}$ \\
         \hline
         DMN-VIS (4-6) & 0.042 & 0.11 & $<10^{-60}$ & 0.035 & 0.11 & $<10^{-52}$ \\
         \hline
         VIS-VIS (6-6) & -0.014 & 0.10 & $<10^{-7}$ & -0.009 & 0.11 & $<0.002$ \\
         \hline
         Total FC & 0.035 & 0.09 & $<10^{-64}$ & 0.031 & 0.087 & $<10^{-62}$ \\
         \hline
    \end{tabular}
    \end{center}
    \label{tab:cross}
\end{table}

In total, there are 9,387 older males ($>65$ years old), 2,425 younger males ($<55$ years old), 8,728 older females ($>65$ years old), and 3,132 younger females ($<55$ years old) in the UKB cross-sectional cohort.

\subsection{Comparison with FC Differences Between Scanner Tasks in the PNC Dataset}
\label{sec:pnc}

\begin{figure}
    \centering
    \hspace*{-2cm}\includegraphics[width=20cm]{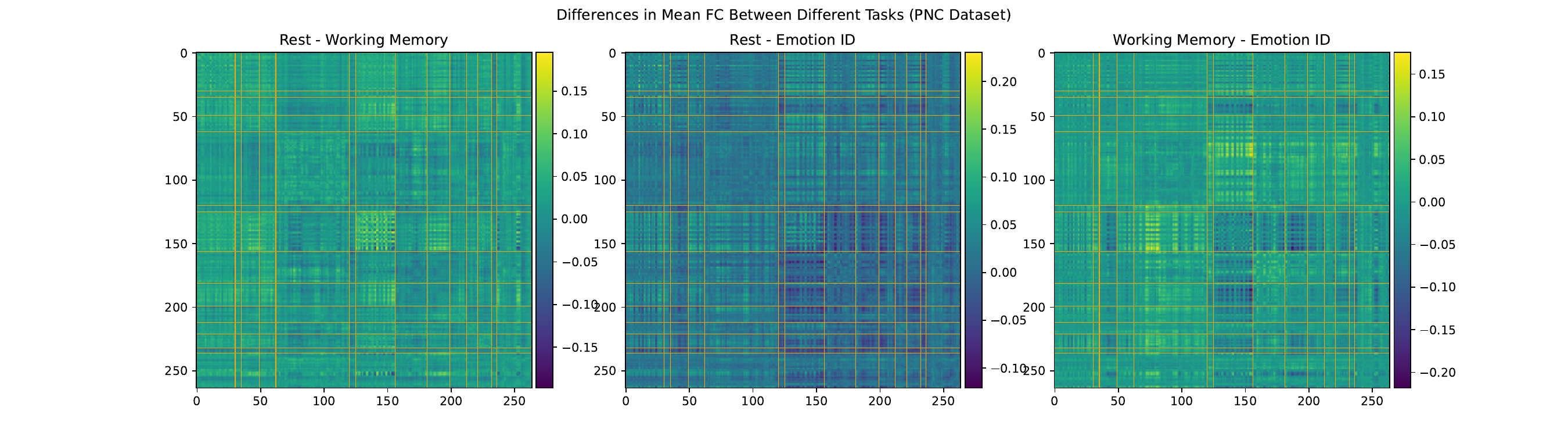}
    \caption{Average differences in FC between three scanner tasks of 1,345 subjects in the PNC dataset. This is a cross-sectional, not longitudinal, dataset.}
    \label{fig:pnc}
\end{figure}

We consider the possibility that the difference in SMT-VIS connectivity between the two scans of the longitudinal cohort is due to a change in scanner task. In Figure~\ref{fig:pnc}, we show the average FC differences between 3 different tasks in the PNC dataset.\cite{Satterthwaite2014NeuroimagingOT} This dataset contains 1,345 children and young adults having all of three different scanner tasks: resting state, working memory, and emotion identification. The preprocessing and FC creation steps for this dataset have been described elsewhere.\cite{10002422} We note that the VIS-VIS is most different for change in task, but that the SMT-VIS is not qualitatively more different that the rest of FC. Also, the magnitude of change in FC in the PNC dataset between tasks is much larger than in the UKB longitudinal cohort.

\section{Discussion}

Farràs-Permanyer et al. (2019)\cite{Farras-Permanyer2019-fx} find that mean resting state FC may increase throughout the entire brain for the oldest subject ($>80$ years old) group. As shown in Appendix~\ref{sec:adni}, we confirm a small, statistically insignificant increase in total longitudinal FC in the healthy controls of the ADNI dataset,\cite{Petersen2010-lg} another elderly population with multiple longitudinal fMRI scans. In Section~\ref{sec:cross}, we show that there is a large, statistically significant increase in average resting state FC across almost all inter-network connections in the UKB cross-sectional cohort with increased age. This cross-sectional cohort is much larger than the longitudinal cohort we describe in the main part of this paper. The fact that SMT-VIS FC also increases in the cross-sectional cohort, but not disproportionately compared to the rest of FC, raises the possibility of a change in resting state scanner task during the second longitudinal scan. We believe this longitudinal change is not an artifact of our pre-processing methods. Credence should increase in our pre-processing methods since the UKB-provided ICA-based FC and PC is also able to predict longitudinal scan ordering at almost the same level as our Power264-based approach, although the ICA FC and PC matrices are not interpretable.

Many studies have focused on examining connectivity in the DMN associated with aging.\cite{Sala-Llonch2015-ud}\cite{Malagurski2022} These studies find areas of increased connectivity as well as areas of decreased connectivity. There are two problems with such studies. First, they are for the most part cross-sectional and do not follow a single subject across a multi-year period. Second, they mostly use small numbers of subjects, the majority of studies enrolling fewer than 50, making it impossible to identify small effects. On the other hand, one study performed on a cohort of more than 2,000 older subjects in Rotterdam found age-related changes in connectivity to be complicated, drawing no firm conclusions.\cite{Zonneveld2019-hs} We note that the Rotterdam study was not longitudinal but cross-sectional. 

We conjecture the fact that most studies only focus on DMN and report decreased connectivity\cite{Sala-Llonch2015-ud} in aging populations may be related to the large number of ROIs in the DMN and an implicit bias inherent in the word ``connectivity." Naturally, as we reach very old age we expect physical connections to degenerate, not become stronger. In fact, FC is really the synchronization of BOLD signal between regions, and does not imply a direct physical link between regions. Young children are known to have higher average FC than young adults;\cite{Dosenbach2010-ef}\cite{Jolles2011-uz} thus older subjects may been as reverting to a less optimal state as they age.

On the other hand, as we describe in Appendix~\ref{sec:corr}, physical observables such as hand grip strength in the UKB longitudinal cohort are weakly correlated with an increase in FC in SMT-CB and VIS-CB connectivity. Additionally, we find BMI and basal metabolic rate are weakly correlated with the longitudinal increase in SMT-VIS connectivity (see Appendix~\ref{sec:corr}). This may suggest a small health related effect that is found throughout the study cohort and includes male, female, younger, and older subjects. Finally, as discussed in Section~\ref{sec:pnc}, we cannot rule out the possibility that changes in longitudinal FC are caused by a change in scanner task between acquisition of the two longitudinal timepoints.

\begin{figure}
    \centering
    \includegraphics[width=8cm]{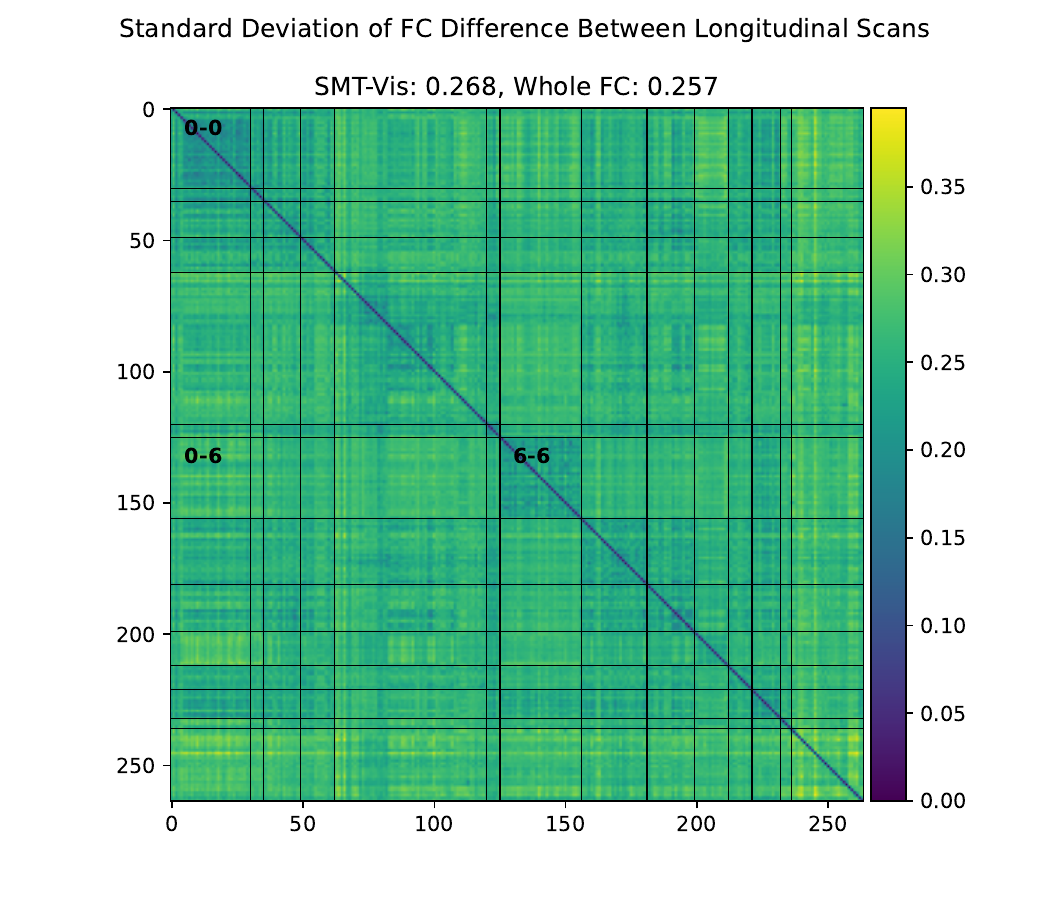}
    \caption{Standard deviation of difference between older scan FC and younger scan FC in the UKB longitudinal cohort. Note that the magnitude of average standard deviation (0.26) of SMT-VIS connectivity change is large compared to mean SMT-VIS connectivity change (0.03). However, we show in Figure~\ref{fig:pred-inter-net} that this connectivity change is highly statistically significant. The smallest average standard deviations are found in SMT-SMT (0.2) and VIS-VIS (0.22) connectivities.}
    \label{fig:std}
\end{figure}

We show in this work that the average connectivity increase in the SMT-VIS connection is small but highly statistically significant. The average change in FC in this connection is only $6.8\%$, corresponding to a mean change $\mu_{\Delta\rho}=+0.03$, compared to a standard deviation of change from subject to subject of $\sigma_{\Delta\rho}=0.26$ (see Figure~\ref{fig:std}). However, using our longitudinal sample of 2,722 subjects, we find the average SMT-VIS connectivity change from younger scan to older scan is significant as level of $p<10^{-15}$ after Bonferroni correction for multiple comparisons (Figure~\ref{fig:pred-inter-net}). Finding such small effects is helped by the use of large number of subjects and longitudinal data.

\section{Conclusion}

In this work, we pre-process a 2,722 subject longitudinal subset of the UK Biobank dataset and examine FC using the Power264 atlas. We find that in scans taken an average of two years apart, the average functional connectivity between SMT and VIS network regions tends to increase. This occurs in male, female, younger ($<55$ years old), and older ($>65$ years old) subjects. We verify the ability of this average FC increase to predict scan ordering using simple machine learning models. The identification of an increase in connectivity with non-pathological aging, in longitudinal as well as cross-sectional cohorts, and specifically in the SMT-VIS synchronization of BOLD signal, may lead to novel insights about brain function in old age. Additionally, we identify an effect that could possibly show up as a confounder in studies of dementia or neurodegenerative diseases. Nonetheless, we remain open to the idea of a change in resting state scanner task during acquisition of the longitudinal data in the UKB being partly responsible for this effect.

\section*{DISCLOSURES}

The authors have no conflicts of interest to report.

\def\UrlBreaks{\do\/\do-}
\section*{CODE AND DATA AVAILABILITY}

fMRI and phenotype data came from the UK Biobank (application ID 61915), available via application to qualified researchers. Additional data used in Appendix~\ref{sec:adni} came from the Alzheimer's Disease Network Initiative (ADNI), available via application from \url{https://adni.loni.usc.edu/}. Additional neuroimaging data came from the Neurodevelopmental Genomics: Trajectories of Complex Phenotypes database of genotypes and phenotypes repository, dbGaP Study Accession ID phs000607.v3.p2.

All code used in this study is available from GitHub at \url{https://github.com/aorliche/ukb-longitudinal-smt-vis}. We do not have permission to post original subject data; however, it may be obtained via application from the sources listed above.

\acknowledgments 
The authors would like acknowledge the NIH (grants R01 GM109068, R01 MH104680, R01 MH107354, P20 GM103472, R01 EB020407, R01 EB006841, R56 MH124925) and NSF (\#1539067) for partial funding support.

This research was supported in part using high performance computing (HPC) resources and services provided by Information Technology at Tulane University, New Orleans, LA.

\section*{BIOGRAPHY}

\textbf{Anton Orlichenko} received his B.S. in Electrical and Computer Engineering in 2010 from the Illinois Institute of Technology. He is currently a Ph.D. candidate at Tulane University, working on mathematical models of brain function and genetics.

\textbf{Kuan-Jui Su} received his Ph.D. from Tulane University School of Medicine in 2023. He currently works at the Center for Biomedical Informatics and Genomics at Tulane University.

\textbf{Qing Tian} holds an M.S. degree and works with Dr. Deng and Dr. Shen as the database administrator for the Center for Biomedical Informatics and Genomics at Tulane University.

\textbf{Hui Shen} received his B.S. from the University of Science and Technology in China and his Ph.D. from Creighton University. He is currently a geneticist at the Center for Biomedical Informatics and Genomics at Tulane University.

\textbf{Hong-Wen Deng} received his B.S. and M.S. from Peking University. He received a further M.S. and Ph.D. degrees from the University of Oregon. He currently runs the Center for Biomedical Informatics and Genomics at the Tulane University School of Medicine.

\textbf{Yu-Ping Wang} received his B.S. from Tianjin University in 1990, and his M.S. and PhD degrees from Xian Jiaotong University in 1993 and 1996, respectively. He currently runs the Multiscale Bioimaging and Bioinformatics Lab at Tulane University.

\bibliography{report}
\bibliographystyle{spiebib} 

\newpage

\appendix

\section{Longitudinal FC Changes in the ADNI Dataset}
\label{sec:adni}

\begin{figure}
    \centering
    \includegraphics[width=14cm]{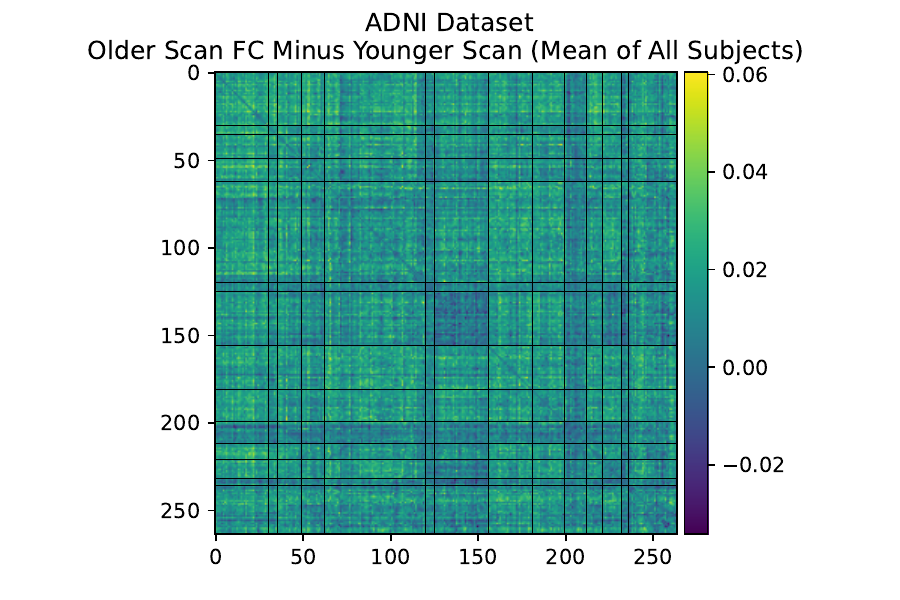}
    \caption{Mean longitudinal changes in FC between first scan and second scan in healthy controls of ADNI dataset.}
    \label{fig:adni-fc}
\end{figure}

We examined the longitudinal change in FC of healthy controls in the Alzheimer's Disease Neuroimaging Initiative (ADNI) dataset\cite{Petersen2010-lg} (age matched subjects who do not develop AD pathology). We used scans taken an average of one year apart. We confirm a small, statistically insignificant increase in total FC but fail to find the same SMT-VIS increase relative to the rest of FC as in the UKB. Statistics are given in Table~\ref{tab:adni} and the average FC change is shown in Figure~\ref{fig:adni-fc}.

\begin{table}[h]
    \centering
    \begin{tabular}{|c|c|c|c|c|}
        \hline
         Average FC Increase & Std Dev FC Change & +Total FC & -Total FC & Total Subjects \\
         \hline
         0.015 & 0.131 & 184 (52.7\%) & 165 & 394 \\
         \hline
    \end{tabular}
    \caption{There is a small but positive change in FC in ADNI healthy controls (subjects who do not go on to develop AD).}
    \label{tab:adni}
\end{table}

\newpage

\section{Correlation of Change in FC with Longitudinal Outcomes in the UKB}
\label{sec:corr}

We identified several correlations between longitudinal change in FC and changes in clinical outcomes associated with the two scan timepoints in the UKB dataset. These are presented below, along with the UKB field identifiers of the outcomes. All p-values are Bonferroni-corrected with $n=105$ multiple comparisons (one for each average inter-network connectivity).

\subsection{SMT Hand, VIS, and CB Connectivity and Grip Strength (f.46.2.0, f.46.3.0, f.47.2.0, f.47.3.0)}

\begin{figure}
    \centering
    \includegraphics[width=14cm]{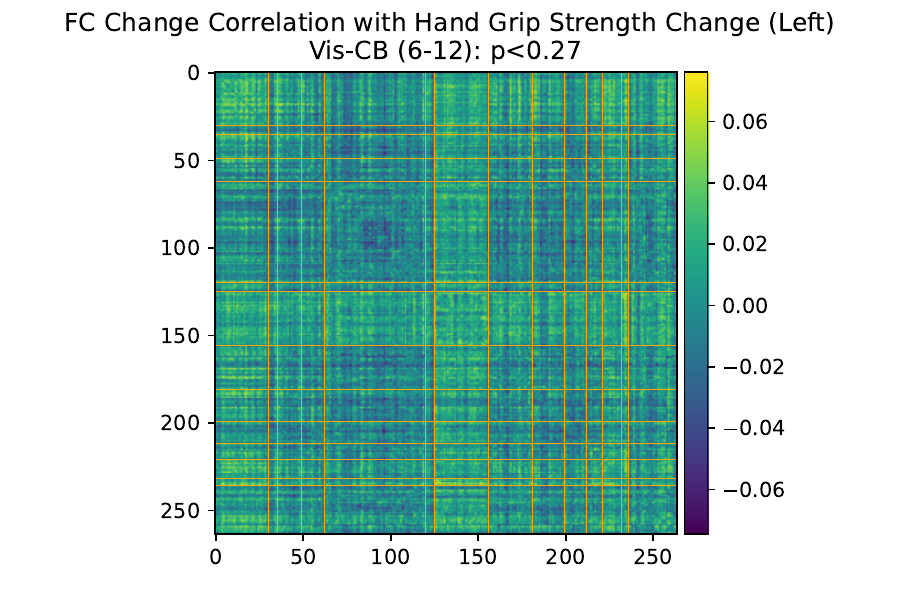}\\
    \includegraphics[width=14cm]{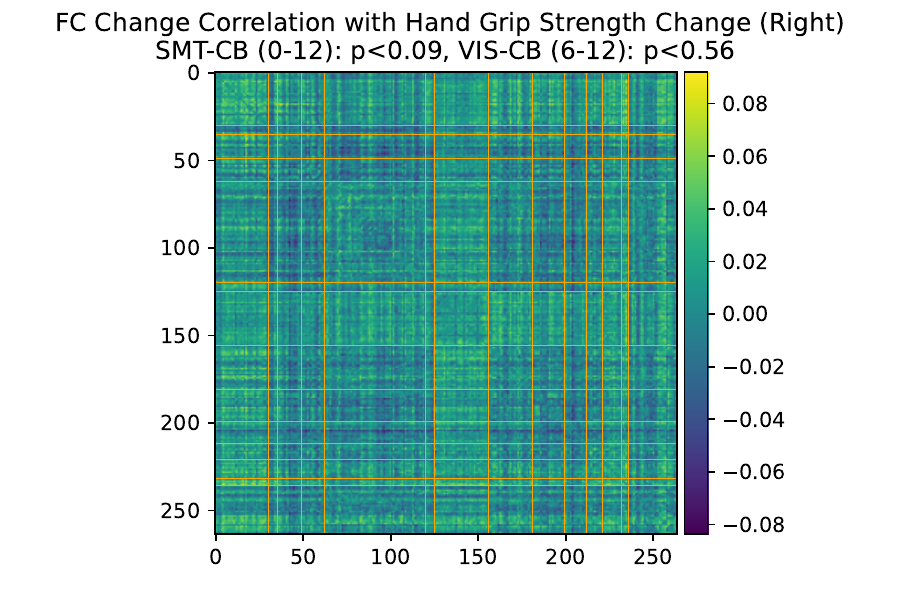}
    \caption{Hand grip strength change association with functional connectivity change in the longitudinal cohort. Bonferroni-corrected p-values.}
    \label{fig:hand-grip}
\end{figure}

We find a marginally significant association between change in hand grip strength and VIS-CB and SMT-CB connectivity change (Figure~\ref{fig:hand-grip}).

\subsection{Body Mass Index and Basal Metabolic Rate (f.23104.2.0, f.23104.3.0, f.23105.2.0, f.23105.3.0)}

\begin{figure}
    \centering
    \includegraphics[width=14cm]{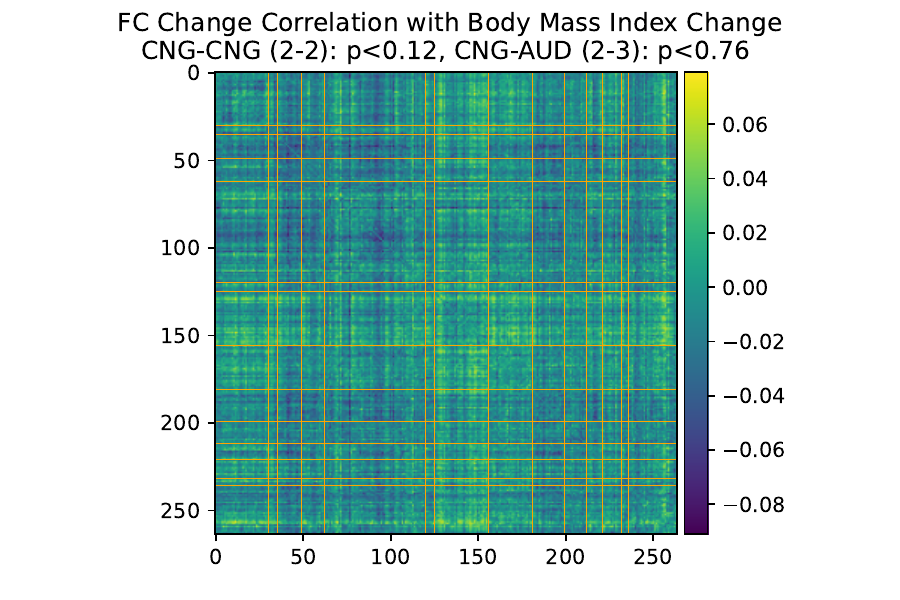}\\
    \includegraphics[width=14cm]{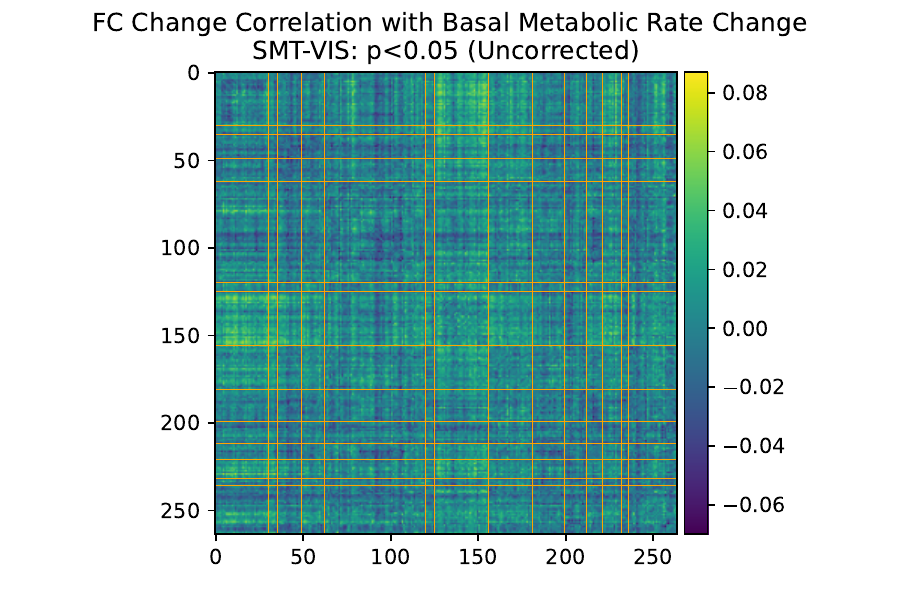}
    \caption{BMI and basal metabolic range change association with functional connectivity change in the longitudinal cohort. Bonferroni-corrected p-values.}
    \label{fig:bmi-smt-vis}
\end{figure}

We find a not statistically significant but suggestive association between BMI and basal metabolic rate change and SMT-VIS connectivity change (Figure~\ref{fig:bmi-smt-vis}).

\end{document}